# Visualizing landscapes of the superconducting gap in heterogeneous superconductor thin films: geometric influences on proximity effects

Jungdae Kim, Victor Chua, Gregory A. Fiete, Hyoungdo Nam, A. H. MacDonald, and Chih-Kang Shih[*]

Department of Physics, The University of Texas at Austin, Austin, Texas 78712, USA

*Corresponding Author: shih@physics.utexas.edu

The proximity effect is a central feature of superconducting junctions as it underlies many important applications in devices and can be exploited in the design of new systems with novel quantum functionality[2-14]. Recently, exotic proximity effects have been observed in various systems, such as superconductor-metallic nanowires[6-8] and graphene-superconductor structures[5]. However, it is still not clear how superconducting order propagates spatially in a *heterogeneous* superconductor system. Here we report intriguing influences of junction geometry on the proximity effect for a 2D heterogeneous superconductor system comprised of 2D superconducting islands on top of a surface metal. Depending on the local geometry, the superconducting gap induced in the surface metal region can either be confined to the boundary of the superconductor, in which the gap decays within a short distance (~ 15 nm), or can be observed nearly uniformly over a distance of many coherence lengths due to non-local proximity effects.

The sample system is comprised of superconducting 2D Pb islands on top of a single atomic layer surface metal, the stripe incommensurate (SIC) phase of Pb overlayer on Si(111)[15-18]. The scanning tunneling microscopy (STM) image shown in Fig. 1 captures a variety of junction configurations. Fig. 1(A) shows an interesting "π" shape Pb island 5 monolayers (ML) thick on top of the striped incommensurate (SIC) surface. Previous scanning tunneling spectroscopy (STS) studies have shown that the SIC phase is superconducting with $T_{C\_SIC}$ = ~ 1.8 K[19] while the 2D Pb islands have a $T_C$ around 6 K[20], although the actual $T_C$ value also depends on the lateral size as well as its thickness[21]. At 4.3 K the SIC template is in the normal state. At locations far from the Pb-islands, the tunneling spectrum exhibits no gap (spectrum #2 in Fig. 1(B)), while the spectrum acquired at the 2D Pb island shows a clear superconducting gap (spectrum #3). In the SIC region near the 2D Pb island, a superconducting gap can also be observed (spectrum #1), indicative of a proximity effect. To address the spatial dependence, we performed spectroscopic mapping over the same area, whose differential conductance at zero bias (zero-bias-conductance (ZBC)) is shown in Fig. 1(C). Since the ZBC directly correlates with the size of the tunneling gap (the *smaller* the value of ZBC, the larger the tunneling gap) the landscape of ZBC is a good representation of the landscape of the superconducting gap.

The ZBC image reveals a rich landscape. In some regions (e.g., region α), the induced superconducting gap decays very quickly within a very short distance from the SIC/superconductor (S) interface, while in region β where the SIC wetting layer is surrounded by Pb islands from both sides, the induced superconducting gap is quite uniform. Similarly, in region γ where the SIC region is near the "recess" of the Pb island, the induced superconducting gap propagates quite extensively. Fig. 1(D) illustrates the spatial dependence of the tunneling spectra along the arrow (starting at position (1) and ending at point (2), as shown in Fig.1(A)), displaying a deep and uniform superconducting gap in the channel region which then decays quickly beyond the region marked by the dashed line.

These diverse superconductor configurations are investigated further according to the local junction geometry. Fig. 2 is the case for a simple S – Normal metal (N) junction with the normal metal region being the SIC template. The STM image and corresponding ZBC image are in Fig. 2(A) and (B), respectively. The short ranged proximity effect at the SIC surface near the junction interface is clearly evident from the ZBC image and this can be confirmed in the color scale plot of dI/dV in Fig. 2(C) taken along the arrow labeled in Fig. 2(A). Shown in 2(D) are the line profiles of ZBC across the junction and the profile of superconducting gap along the same line ($\Delta$ vs. $x$). Note that the ZBC scale is upside-down to be consistent with the strength of the gap. The gap value is obtained by fitting the normalized dI/dV spectra with the BCS density of state (examples shown in inset)[1]. More detailed descriptions of the gap value deduction can be found in Ref. 20 and 21. From the ZBC vs. $x$, as well as $\Delta$ vs. $x$ line profiles, a proximity length of about 12 nm is deduced. We have studied many simple S-SIC junctions and found a consistent behavior: In all cases, the induced gap decays very quickly from the interface with a proximity length of 15±5 nm. In addition, the gap value in isolated Pb islands is always very uniform (to within 5%). Nevertheless, a hint of very small decay (< 5%) can be identified occasionally (as shown in Fig. 2(D)). Except for the fact that $\Delta$ is very uniform up to the very edge of Pb islands, this observation is qualitatively consistent with a phenomenological description by Tinkham[22] as summarized in Fig. 2(E). In this picture, the decay length in the metal, $\xi_M$, is predominantly a property of the metal in proximity contact with S, $\xi_S$ is the effective superconducting coherence length, and *b* is the extrapolation length which depends on the nature of the junction. In our case, $\xi_M$ is about 15 nm while *b* is very large (in the limit of $b \rightarrow \infty$, the Δ in S would be a constant). $\xi_M$ depends on the mean free path of the SIC surface metal which inevitably contains crystallographic defects. A large *b* value may be a result of the fact that most of the S island edge interfaces with the vacuum and the S-SIC interface exists *only along the bottom edge*, which is a line contact a single atom thick. The simple S-S heterojunction follows the conventional picture where the order parameter changes continuously across the interface with a characteristic length scale of about 25 nm on either side of the junction (Fig. 2(F)). Interestingly, this characteristic length scale turns out to be very similar to the coherence length deduced from magnetic susceptibility measurements on Ge-capped thin Pb-films[25].

These conventional behaviors of proximity effects at a simple single-sided S-N or S-S junction change dramatically in a more complicated junction geometry. Fig. 3(A) shows two separated Pb islands with the SIC layer in between, creating a variety of junction geometries. For example, along arrows 1 and 2 are S-N-S heterojunctions with different widths of SIC (N) in between. Also marked with arrow 3 is a simple S-N junction. We map out the induced superconducting gap as a function of distance in the SIC regions along these three arrows using STS. The resulting Δ vs. *x* profiles at two different temperatures are plotted in Fig. 3(B). Also shown in Fig. 3(C) and Fig. 3(D) are the color-scale representation of the dI/dV maps along the two S-N-S double junctions. The spatial mapping of spectra along the arrow (1) revealed a *surprisingly uniform* and well-developed superconducting gap, in contrast to a quick decay of induced gap being observed for the simple S-N junction of arrow (3). Similarly, along arrow (2) the induced gap is also relatively uniform, albeit with a smaller gap value than that along arrow (1). Most surprisingly, this relatively uniform gap value extends over a length scale which is about 4 times the characteristic decay length ($\xi_M \sim 15$ nm), indicating that the S-N-S junction is fundamentally different from the one-sided S-N junction. A uniform $\Delta$ vs. $x$ characteristics is not consistent with a simple superposition of two single-sided S-N junctions. We observe this *non-local* proximity effect in all S-N-S double junctions with widths of up to 90 nm. Considering that the Pb island and the SIC wetting layer is connected only through a line contact of one atom thick, this extended proximity effect is quite astonishing.

This enhanced proximity effect in the double-sided junction is also observed in the surrounded junction where a hole (or a few holes) exposes the SIC in a Pb island (see Fig. 4(A)). In such "surrounded" junctions, we observe a robust superconducting gap in the SIC for holes with an effective diameter of up to 60 nm. The ratio between the gap value measured in the hole and that measured in the surrounding Pb island remains relatively constant over a wide range of temperatures. Fitting $\Delta$ vs. $T$ with the BCS model, one finds that the data at different spatial locations (Pb island, different holes) converge to the same $T_C$ due to this constant gap ratio. Since the superconducting gap observed in the SIC is induced by the proximity effect, the convergence to the same $T_C$ may not be a surprise.

The ratio between the gaps measured on the surrounding Pb islands and at the center of the SIC structures shows a systematic trend with respect to the effective hole diameter of the surrounded junction (larger diameter, smaller gap), as shown in Fig.4(B). Similar trends can be observed for the two-sided S-N-S junction albeit with a steeper slope by about a factor of two.

These rich phenomena exemplify the intriguing manifestations of junction geometry on the proximity effect. At a simple S-N junction, the induced gap exists very locally, being confined within a length scale of $\xi_M \sim 15$ nm. On the other hand, if there is another Pb island nearby, then the induced gap persists to a length scale many times $\xi_M$. Such a behavior is in stark contrast to the conventional picture predicted by solving the GL equations. We note that in a recent investigation of 1D S-N-S junction, minigaps exist in the normal metal nanowires over an extended region[8]. Moreover, the result is quantitatively consistent with the solutions of the non-linear Usadel equation[10]. Applying the Usadel equations using the parameters determined experimentally in the simple S-N junction (see supplemental information)[10-12], one finds the solution reproduces a "uniform" value of minigap in the normal metal region, whose value decreases as the width of normal region increases. However, we also noted clear numerical differences associated with dimensionality. Experimentally, the gap decrease linearly with L which deviates from the 1D Usadel solutions (see Fig.4(B)). In addition, there exist other profound geometric influences (such as curvature, shape of the junction, and the effect of a surrounded junction) on the proximity effect in complex superconductor – metal junction geometries. These complicated superconducting gap landscapes can be qualitatively understood, however, by considering the possible multiple Andreev reflections. For example, in Fig 1(C) one can see that in region $\alpha$ where the SC island bends into the SIC, a very short decay length is observed. One the other hand, in region $\gamma$ where the SIC bends into the S island, the induced superconducting gap exists over a more extended range since more semi-classical Andreev reflections are allowed by the local geometry. The multiple reflections also explain why in the surrounded junction (hole in SC island) the induced superconducting gap is so much more robust than that in the SC-SIC-SC junction, as illustrated Fig. 4(A) and (B).

In summary, our experiments have given us unprecedented spatial detail on the superconducting proximity effects in spatially inhomogeneous superconducting-normal systems. We have found that certain island geometries provide "conventional" proximity effects, while other geometries reveal a wealth of information about how the "shape" of a superconductor and the junction influence superconductivity. In particular, we find a type of "giant" proximity effect in a very conventional superconducting system that can be engineered through geometry alone. This observation may enable new functionality in superconductor-based devices, and may portend even more intriguing phenomena in spatially inhomogeneous superconducting-normal systems composed of "exotic" superconductors.

**Acknowledgements**: We are grateful to Leonid Glazman and Alex Kamenev for discussions, and to grants ARO W911NF-09-1-0527 and NSF DMR-0955778.

## Methods

**Sample Preparation**

The experiments were conducted in a home-built low temperature STM system with an *in-situ* sample preparation chamber. A Pb-Si reconstructed surface of the striped incommensurate (SIC) phase was prepared by deposition of ~ 1ML of Pb onto the Si(111) 7×7 surface at room temperature, followed by sample annealing at 400 ~ 450 °C for 4 min to form the surface template (see supplementary Fig. S1). In order to get 2D islands, Pb was deposited on the template at ~ 200 K with a deposition rate of 0.5ML per minute. Before transferring *in-situ* to the LT-STM stage, the sample is annealed shortly at ~ 200 K for 3 minutes. Pb 2D islands on top of SIC surfaces with a various shapes and thicknesses make an ideal system to investigate the influence of geometry on proximity effects.

**Scanning Tunneling Spectroscopy**

Electro-chemically etched tungsten tips treated with *in-situ* e-beam cleaning were used for all measurements. All differential conductance spectra were taken with the same tunneling parameter with the junction stabilized at $V_s$ = 20 mV and $I_t$ =30 pA tunneling current. To eliminate possible piezo creeping and thermal drift, piezo scanner (an STM tip) is stabilized at the required area for more than 8 hours at 4.3 K prior to each zero-bias-conductance (ZBC) measurement. The same tunneling parameter (stabilized at 17 mV with 20 pA tunneling current) was used for all ZBC measurement.

# Figure legends

**Figure 1**

(A) STM topography image of a "π" shape Pb island sitting on top of SIC surface (sample bias $V_s$ = 0.3 V, tunneling current $I_t$ = 10 pA). (B) Differential conductance spectra at 4.3 K measured from different locations labeled in (A). (C) Normalized ZBC image measured at 4.3 K for the same area of (A) which visualizes the dramatic transition of induced superconducting gap on SIC surface between inside and outside of "π" shape as the color contrast reflects the variation of local superconductivity. (D) The spatial dependence of superconducting gap spectra at 4.3 K measured along the arrow in (A). The white dashed line represents the border of confined geometry in (A).

**Figure 2**

(A) STM topography image showing a simple junction of SIC surface – Pb island (sample bias $V_s$ = 0.3 V, tunneling current $I_t$ = 10 pA). (B) Normalized ZBC image measured at 4.3 K for the same area of (A). (C) The spatial dependence of superconducting gap spectra at 4.3 K measured along the arrow in (A). (D) ZBC profile and the fitted superconducting gap profile extracted from (C) which shows the variation of superconducting order across the physical boundary. Note that the ZBC profile is plotted with high value (small gap) pointing down to be qualitatively consistent with the profile of the gap value. (E) Schematic profile describing the qualitative behavior of the order parameter/pair correlations across a conventional S-N junction. (F) The ZBC profile across a superconductor-superconductor (S-S) junction formed by a 3ML Pb islands in connection to a larger 2ML Pb island (the inset is the corresponding 50 nm x 50 nm STM image),

**Figure 3**

(A) STM topography image of two 5ML Pb islands which provide confined geometry for the SIC area marked as arrow (1) and (2) (sample bias $V_s = 0.3$ V, tunneling current $I_t = 10$ pA). (B) The extracted energy gap $\Delta$ values as a function of distance along the arrows in (A). The length of arrow (1), (2), and (3) is around 25 nm, 55 nm, and 40 nm, respectively. The spatial dependence of superconducting gap spectra at 4.3 K measured along (C) the arrow (1) and (D) the arrow (3).

**Figure 4**

(A) The superconducting energy gap as a function of temperatures, $\Delta(T)$, measured at different regions of the corresponding STM image in the inset. The STM image shows an island containing 5 ML and 4 ML regions as well as two holes (H#1 and H#2) in the 5ML region exposing the SIC wetting layer. Also shown are the three fitting curves for these $\Delta(T)$ using a BCS gap equation. Note that despite a different gap value, all three curves fall into the same $T_C$ of ~ 5.7 ± 0.1 K. In the inset, the hole size of H#1 and H#2 are around 19 nm and 35 nm, respectively. (B) The gap ratio of $\Delta_{SIC}$ to $\Delta_{Pb}$ at 4.3 K was measured as a function of the junction distance, d, for 1D and 2D confined heterojunctions described by schematics in the right side.

# Figure 1

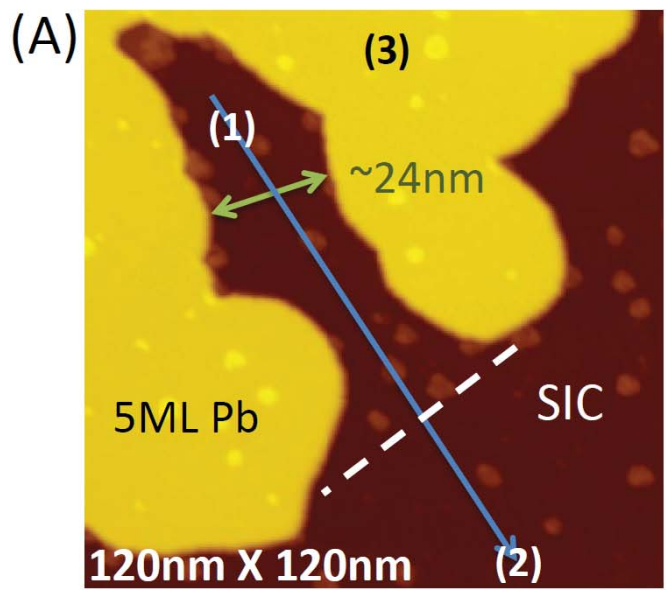
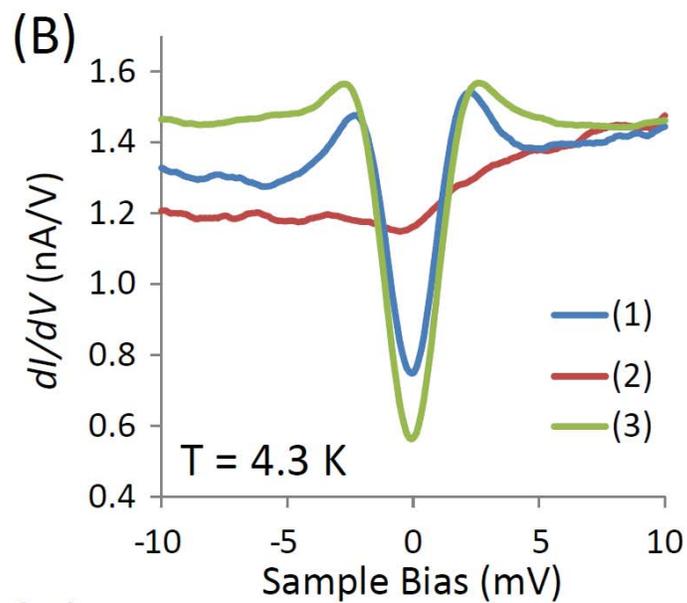
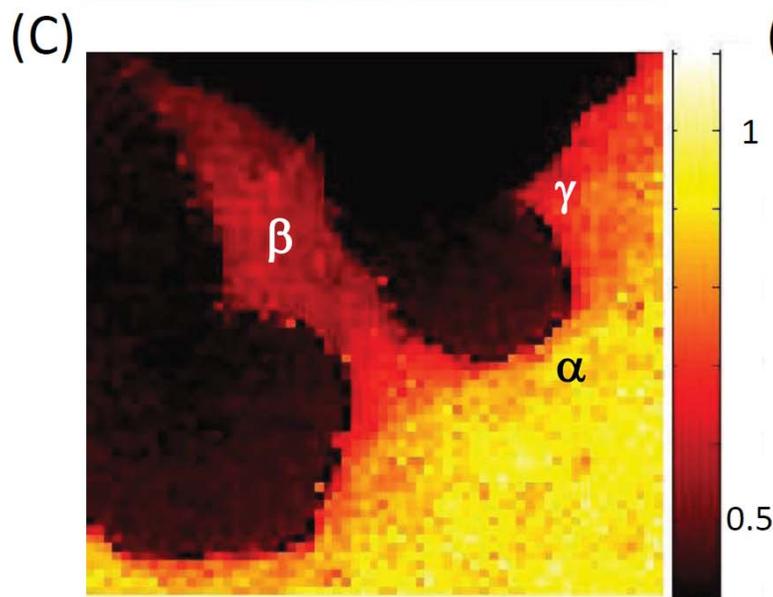
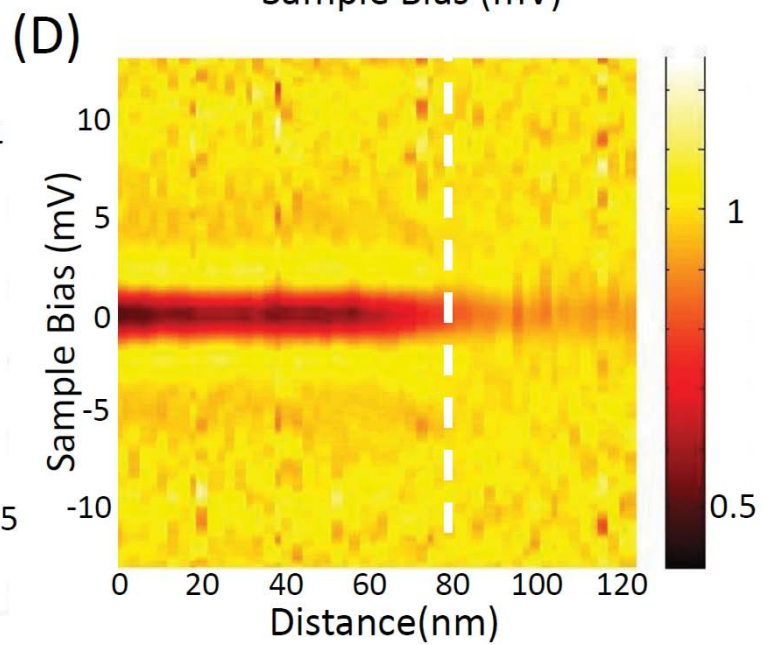

**Figure 2**

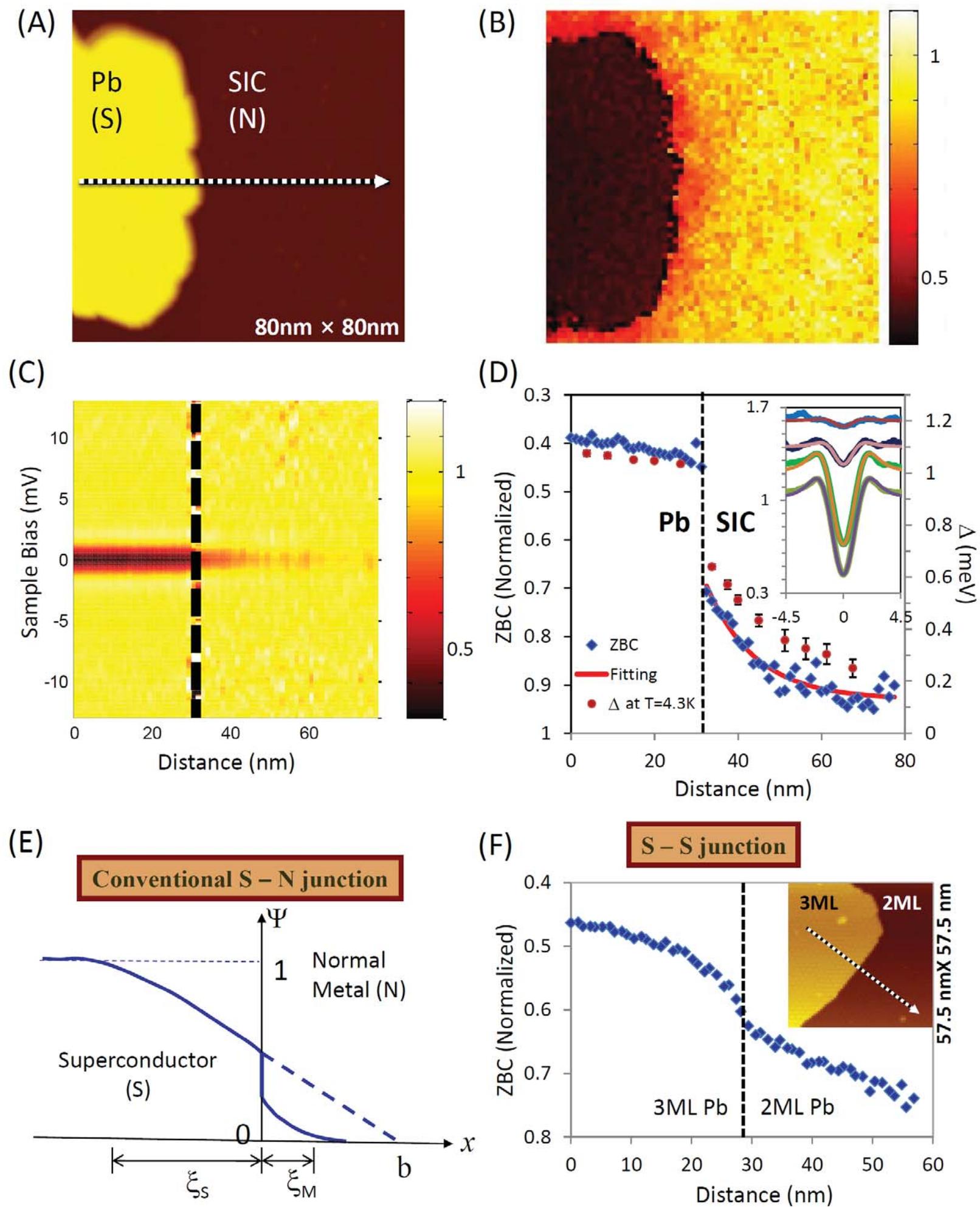

# Figure 3

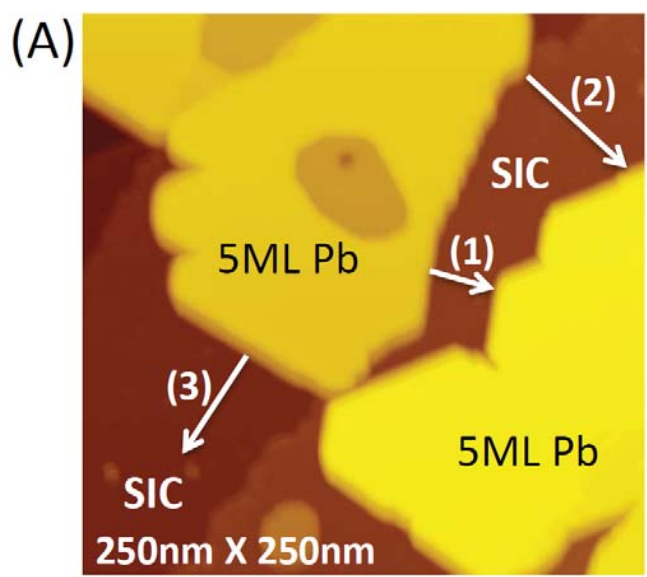
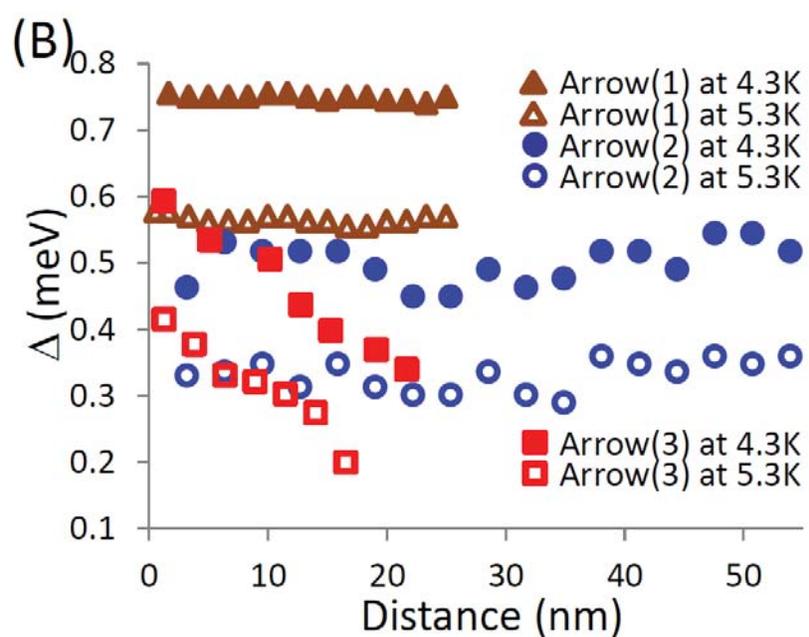
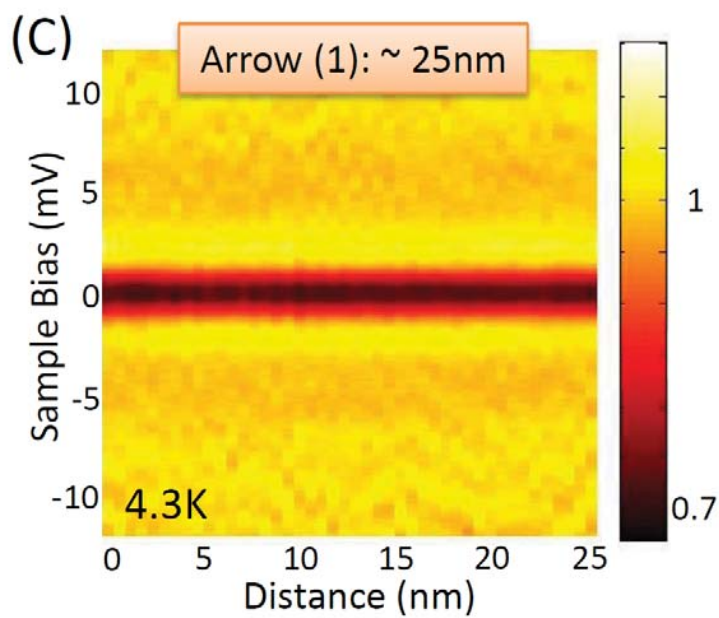
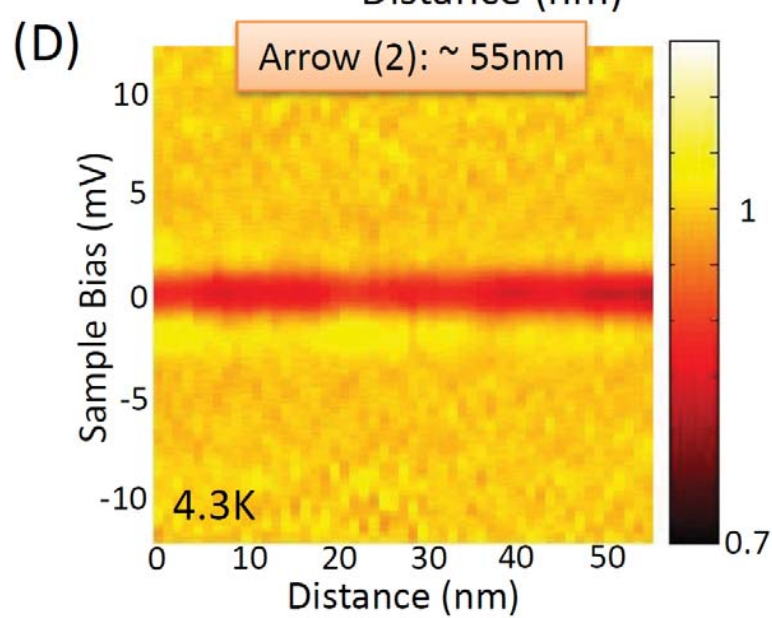

# Figure 4

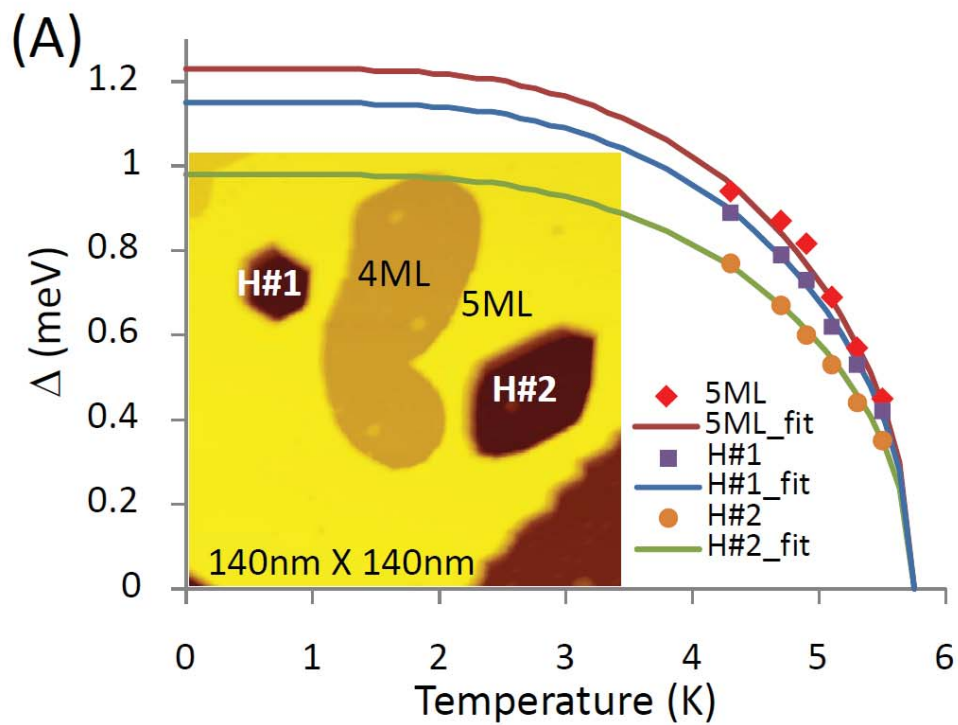

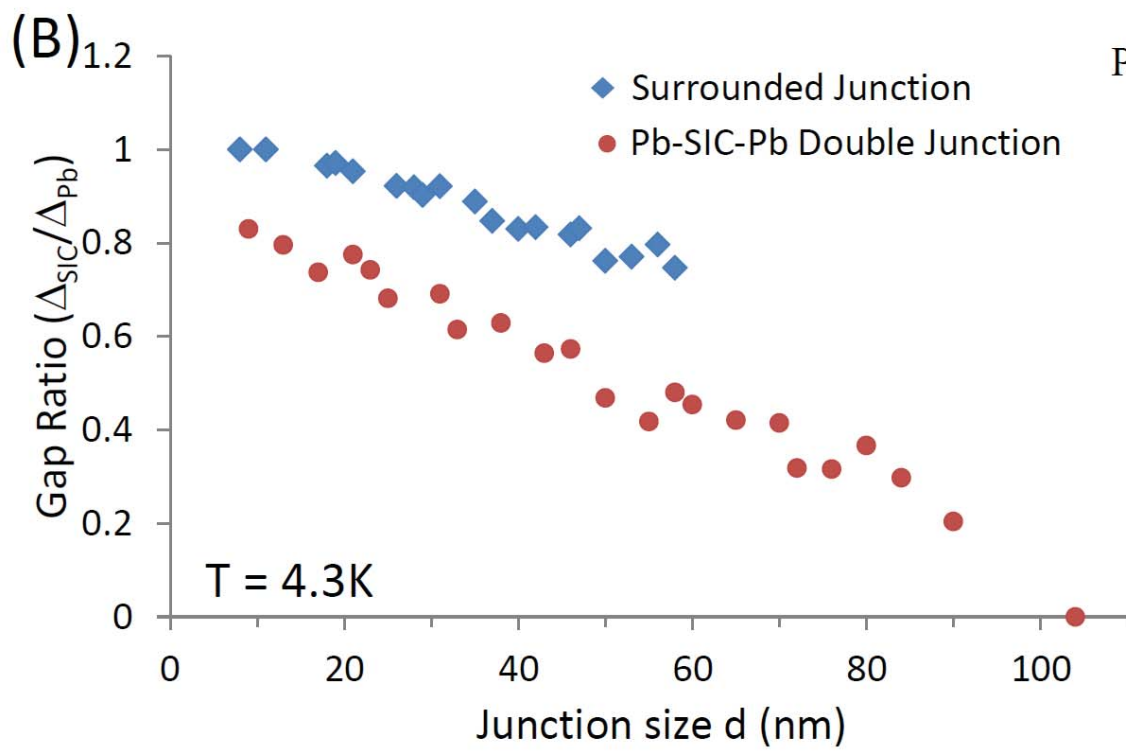

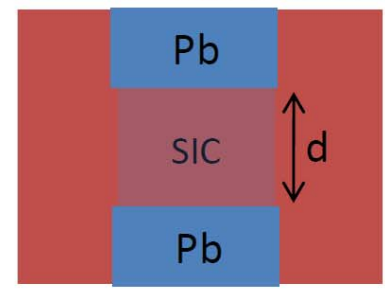

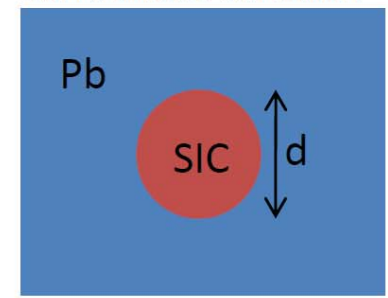

# Supplementary Information for Visualizing landscapes of the superconducting gap in heterogeneous superconductor thin films: geometric influences on proximity effects

Jungdae Kim, Victor Chua, Gregory A. Fiete, Hyoungdo Nam, A. H. MacDonald, and

Chih-Kang Shih*

Department of Physics, The University of Texas at Austin, Austin, Texas 78712, USA

*correspondence to: shih@physics.utexas.edu

**Influence of Pb island geometry on proximity effects**

In addition to Fig. 1, we provide here more examples showing the geometric influence of Pb islands on proximity effects. Although Pb island shapes can be varied as found in Fig. S2(A,C), there is a universal rule on determining the strength and spatial dependence of induced superconducting gaps on the normal state SIC surfaces; i.e. lateral proximity effects. Whenever Pb islands form confined geometries with respect to SIC surfaces, the significant enhancement of proximity effects can be observed as shown in the marked areas of the ZBC images in Fig. S2(B,D).

**The Usadel Equation and Its Solution at T=0**

The starting point for describing the local tunneling density of states is the one dimensional retarded Usadel equation recast in a $\vartheta$ parameterization for the specific case where there is zero phase difference between superconducting order parameters at the boundaries [5, 2],

$$\frac{\hbar D}{2} \frac{\partial^2 \vartheta(x,E)}{\partial x^2} + iE \sinh \vartheta(x,E) - i\Delta(x) \cosh \vartheta(x,E) = 0 \quad (1)$$

where E is the energy, $\vartheta(x,E)$ parameterizes the local superconducting order parameter, $\Delta(x)$ is the position-dependent gap function, and D is the diffusion constant in the SIC. We also make the approximation that the gap function $\Delta$ vanishes in the SIC but remains constant and uniform in the superconducting islands, $\Delta(x,E) = \Delta$. Making the change of variable $\vartheta(x,E) = \frac{i\pi}{2} - \theta(x,E)$ the boundary value problem becomes,

$$\frac{\hbar D}{2}\frac{\partial^2 \theta(x,E)}{\partial x^2} + E\cosh\theta(x,E) = 0, \qquad (2)$$

$$\theta(\pm L/2, E) = i\frac{\pi}{2} - \tanh^{-1}\left(\frac{\Delta}{E+i0^+}\right), \qquad (3)$$

for the one-dimensional region $-\frac{L}{2} < x < \frac{L}{2}$ and $-\infty < E < \infty$, where $L$ is length of the SIC bounded by superconducting islands. The local tunneling density of states (LDOS) at $x$ and energy E is related to $\theta$ by [5, 2],

$$\text{LDOS}(x,E) = N_0 \text{Im}[\sinh\theta(x,E)], \qquad (4)$$

where $N_0$ is the density of states in the normal metal. The boundary value problem (BVP) (2) is then numerically solved by discretizing the differential equation in $x$-space and employing the relaxation method with a simple initial guess [6, 3]. A hundred points were typically used to discretize $[-L/2, L/2]$. With lengths measured in units of coherence length $\xi = (\nabla D/\Delta)^{1/2}$ and energies in meV, (2) was then solved for various values of $E$, below and above $\Delta$. $\text{LDOS}(0, E) \propto \text{Im}[\sinh\theta(0,E)]$ was then used for comparison with experiment. Shown in Fig. S3 is a solution to (2).

**Determining $\Delta$ and $\xi$ for $T = 4.3$K**

Measurements were taken at $T = 4.3$K which has the effect of thermally broadening all tunneling measurements which we take into account by convoluting with the first derivative of the Fermi function, $-\partial_E n_F(E-\mu)$ [4]. In addition, Gaussian broadening was included to account for any additional experimental noise. Deep inside the superconducting region, the BCS expression for the density of states may be used to described the observed the tunneling density of states,

$$\text{LDOS}_{BCS}(E) = \frac{N_0|E|}{(E^2 - \Delta^2)^{1/2}}. \qquad (5)$$

Comparisons of measurements in the superconducting (SC) Pb islands (of similar thickness to the regions of interest) at 4.3K to the BCS expression for the tunneling density of states determines $\Delta$ to be ~1.1meV at noise level $\sigma = 0.81$meV. We take $\Delta = 1.1$meV in all subsequent calculations and the same level of noise.

To determine the coherence length $\xi$, a fit was made to data from the SIC in the vicinity of a single SC island, or SC-normal (SN) geometry. This SN geometry in one dimension has a known solution to the Usadel equation [1] given as follows

$$\theta(E,x) = \frac{i\pi}{2} - 4\tanh^{-1}\left[\tanh\left[\frac{\theta_{BCS}}{4}\right]\exp\left[-\left(\frac{-2iE}{\Delta}\right)^{1/2}\left(\frac{x}{\xi}\right)\right]\right], \; x > 0, \qquad (6)$$

$$\theta_{BCS} = \tanh^{-1}\left(\frac{\Delta}{E+i0^+}\right), \qquad (7)$$

where $x = 0$ is the boundary with the SC. In equation (4), $x = 0$ corresponds to the center of the SNS junction. From this exact expression (6), the zero energy or zero bias, $E = 0$ density of states is then compared to zero bias tunneling data at varying distances from the superconductor. Again before comparison with data, all numerical curves are thermal and then Gaussian broadened as was discussed above. Fig. S4 is a fitted plot against the zero bias conductance data which yields a coherence length of $\xi = 22$nm.

**SIC surrounded by Superconductors: Experiment and Theory**

We use the parameters of the previous section to compare theory and measurements taken from a SIC bounded by two superconducting islands. An interesting observation for the S-N-S (Pb-SIC-Pb) double junction is the difference in dI/dV spectral line-shape between those acquired at the Pb islands and those acquired at the SIC as shown in Fig. S5(A). While the spectra acquired in the SIC show a shallower ZBC and a narrower separation between the coherent peaks (both are consequence of a smaller gap), the amplitude of the coherent peaks turns out to be quite strong with a line shape that differs from the BCS prediction as shown in Fig. S5(B).

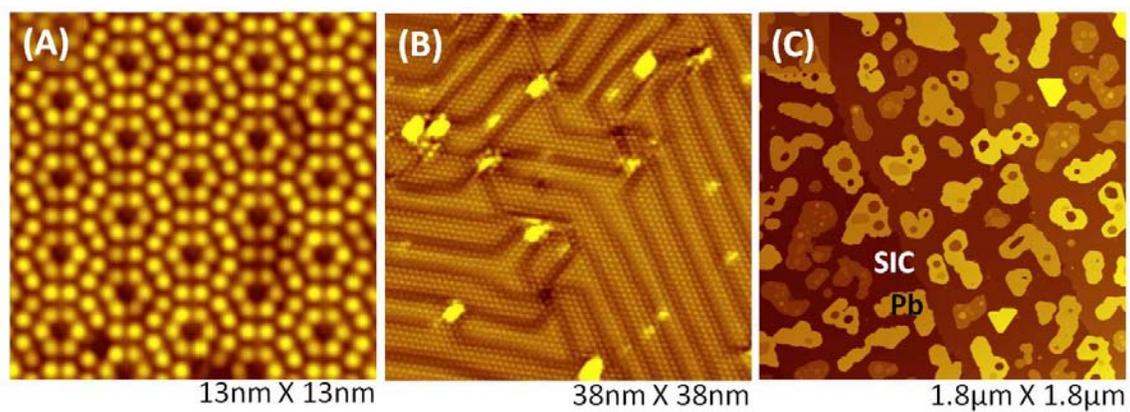

**Fig. S1**
STM topography images of (A) Si(111) 7×7 surface, (B) Striped incommensurate (SIC) phase of Pb on Si(111), and (C) Pb 2D islands on top of SIC surface.

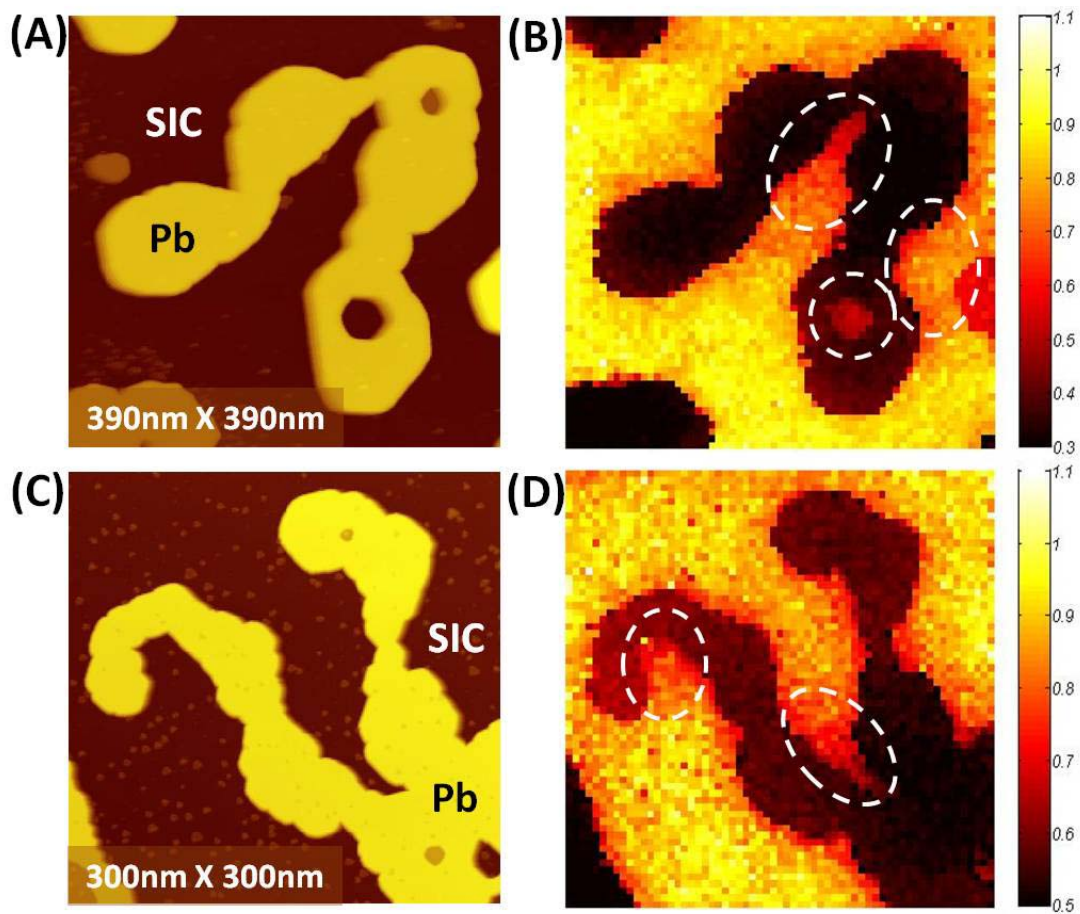

**Fig. S2**

(A, C) STM topography images of Pb islands forming confined geometry and (B, D) corresponding ZBC images. The circled areas in (B, D) show dramatic enhancement of proximity effects due to the confined shape of nearby Pb islands.

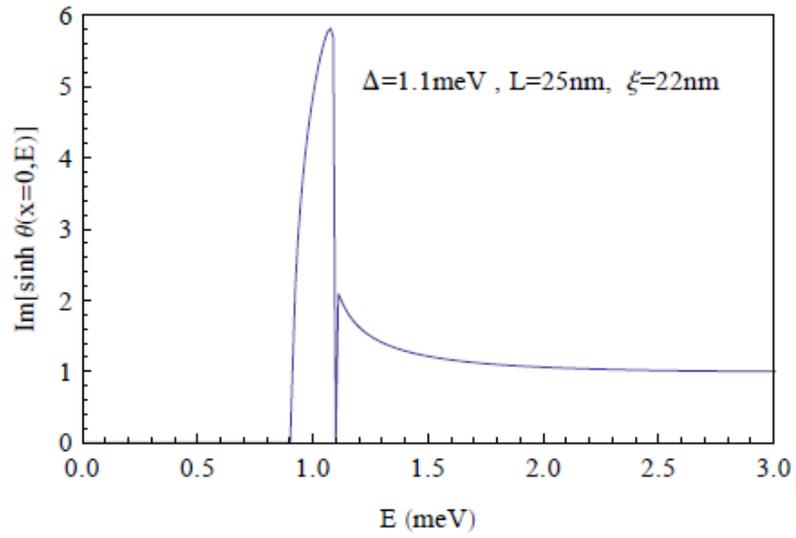

**Fig. S3**
The theoretical local density of states at $x = 0$ at $T = 0$ for a Superconducting-Normal-Superconducting (SNS) arrangement with $\Delta = 1.1$ meV, $L = 25$nm and $\xi = 22$nm.

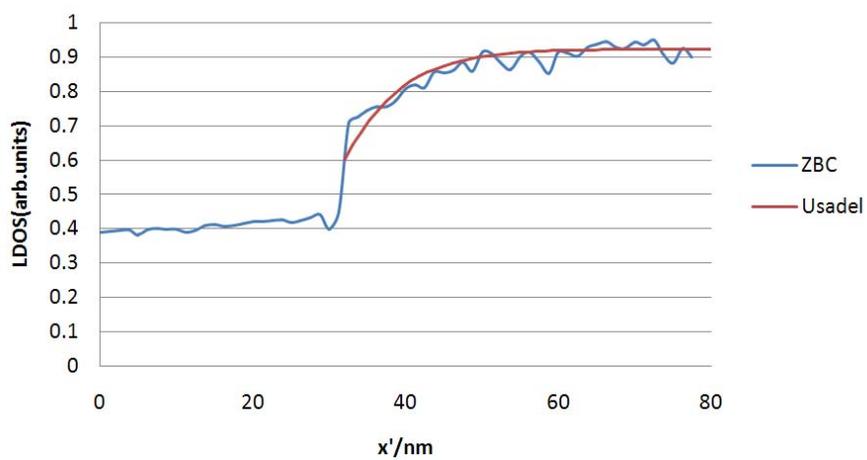

**Fig. S4**
The zero bias conductance data (ZBC) shown in blue and the Usadel prediction of equation (6) shown in red. The interface where the edge of the Pb island lies at is $x' \approx$ 32nm. This fit yields $\xi = 22$nm.

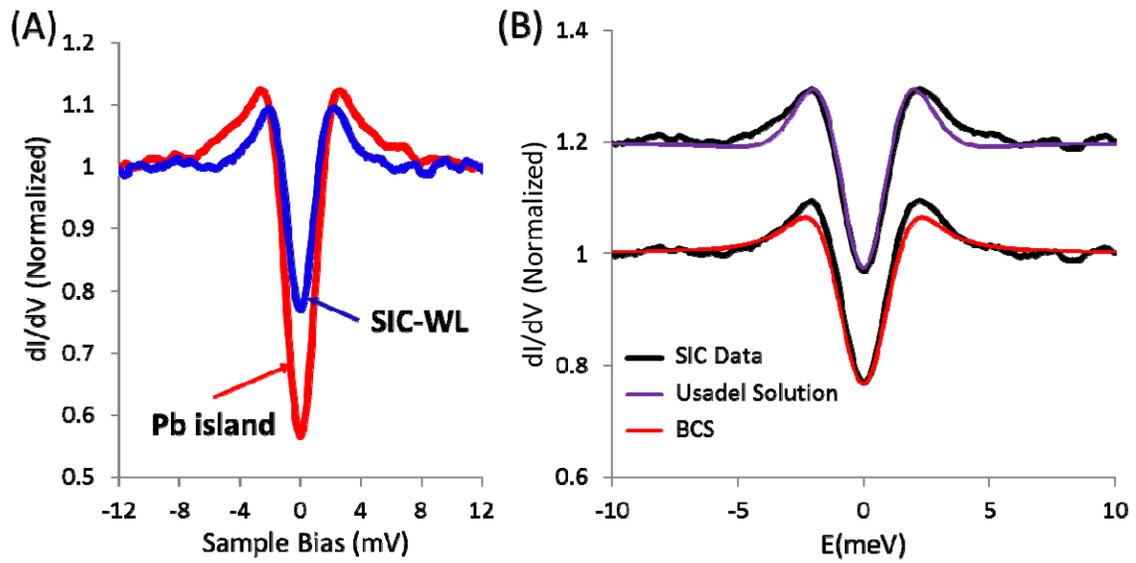

**Fig. S5**
Comparison of the measured data and the Usadel prediction for the S-N-S geometry. (A) Individual tunneling spectra acquired at the 5ML Pb island and at the SIC wetting layer between two Pb islands. (B) Comparison between the spectral line-shape fitted with a BCS like DOS (bottom) and that fitted with the solution of the Usadel equation (top). The top spectrum is offset by 0.2 for clarity.